\documentclass{acm_proc_article-sp}

\usepackage{caption}

\usepackage{amssymb}
\usepackage{graphicx}
\usepackage{verbatim}
\usepackage{amsmath}
\usepackage{url}
\usepackage{cite}
\usepackage{listings}
\usepackage{algorithm}
\usepackage{algpseudocode}
\usepackage{bm}
\usepackage{array}
\usepackage{rotating}
\urldef{\mailsa}\path|{alfred.hofmann, ursula.barth, ingrid.haas, frank.holzwarth,|
\urldef{\mailsb}\path|anna.kramer, leonie.kunz, christine.reiss, nicole.sator,|
\urldef{\mailsc}\path|erika.siebert-cole, peter.strasser, lncs}@springer.com|

\newcommand{\flow}{\mathsf{flow}}
\newcommand{\jump}{\mathsf{jump}}
\newcommand{\inv}{\mathsf{inv}}
\newcommand{\init}{\mathsf{init}}

\newcommand{\reach}{\mathsf{Reach}}
\newcommand{\goal}{\mathsf{goal}}

\newcommand{\lrf}{\mathcal{L}_{\mathbb{R}_{\mathcal{F}}}}
\newcommand{\citep}{\cite}
\newcommand{\hide}[1]{}

\newcommand{\ie}{{\em i.e.}}
\newcommand{\enforce}{\mathsf{enforce}}

\newenvironment{definition}[1][Definition]{\begin{trivlist}
\item[\hskip \labelsep {\bfseries #1}]}{\end{trivlist}}

\begin{document}

\title{SReach: A Bounded Model Checker for Stochastic\\ Hybrid Systems}
%
%
%
%
%

\numberofauthors{5} 

\author{
\alignauthor
Qinsi Wang\\
       \affaddr{Computer Science Department}\\
       \affaddr{Carnegie Mellon University}\\
       \email{qinsiw@cs.cmu.edu}
\alignauthor
Paolo Zuliani\\
       \affaddr{School of Computing Science}\\
       \affaddr{Newcastle University}\\
       \email{paolo.zuliani@ncl.ac.uk}
\alignauthor 
Soonho Kong\\
       \affaddr{Computer Science Department}\\
       \affaddr{Carnegie Mellon University}\\
       \email{soonhok@cs.cmu.edu}
\and  
\alignauthor 
Sicun Gao\\
       \affaddr{CSAIL}\\
       \affaddr{Massachusetts Institute of Technology}\\
       \email{sicung@csail.mit.edu}
\alignauthor 
Edmund M. Clarke\\
       \affaddr{Computer Science Department}\\
       \affaddr{Carnegie Mellon University}\\
       \email{emc@cs.cmu.edu}
}

\maketitle

\begin{abstract}
In this paper we describe a new tool, {\it SReach}, which solves probabilistic bounded reachability problems for two classes of stochastic hybrid systems. The first one is (nonlinear) hybrid automata with parametric uncertainty. The second one is probabilistic hybrid automata with additional randomness for both transition probabilities and variable resets. Standard approaches to reachability problems for linear hybrid systems require numerical solutions for large optimization problems, and become infeasible for systems involving both nonlinear dynamics over the reals and stochasticity. Our approach encodes stochastic information by using random variables, and combines the randomized sampling, a $\delta$-complete decision procedure, and statistical tests. {\it SReach} utilizes the $\delta$-complete decision procedure to solve reachability problems in a sound manner, i.e., it always decides correctly if, for a given assignment to all random variables, the system actually reaches the unsafe region. The statistical tests adapted guarantee arbitrary small error bounds between probabilities estimated by {\it SReach} and real ones. Compared to standard simulation-based methods, our approach supports non-deterministic branching, increases the coverage of simulation, and avoids the zero-crossing problem. We demonstrate our method's feasibility by applying {\it SReach} to three representative biological models and to additional benchmarks for nonlinear hybrid systems with multiple probabilistic system parameters.

\end{abstract}
\section{Introduction} 
Stochastic hybrid systems (SHSs) are dynamical systems exhibiting discrete, continuous, and stochastic dynamics. Due to their generality, SHSs have been widely used in various areas, including cyber-physical systems, financial decision problems, and biological systems \cite{blom2006stochastic, clarke2011statistical}. The popularity of SHSs in real-world applications motivates researchers to put a significant effort into analysis methods for this class of systems. One of the elementary questions for the quantitative analysis of SHSs is the probabilistic reachability problem, i.e., computing the probability of reaching a certain set of states. The set may represent unsafe states which should be avoided or visited only with some small probability, or dually, good states which should be visited frequently. There are two reasons why this kind of problem catches the researchers' attention. One is that most temporal properties can be reduced to reachability problems, considering the very expressive hybrid modeling framework. The other is that probabilistic state reachability is a hard and challenging problem which is undecidable in general. 

\vspace{-.2cm}
To describe stochastic dynamics, uncertainties have been added to hybrid systems in a number of different ways. One of the simplest ways replaces some of the system parameters with random variables, resulting in general hybrid automata (GHAs) with parametric uncertainty. Another approach integrates deterministic flows with probabilistic jumps. When state changes forced by continuous dynamics involve discrete random events, we refer to such systems as probabilistic hybrid automata (PHAs) \cite{sproston2000decidable}. When state changes also involve continuous probabilistic events, we call this kind of models stochastic hybrid automata (SHAs) \cite{franzle2011measurability}. Other models describe randomness by substituting deterministic flows with stochastic ones, such as stochastic differential equations (SDEs) \cite{ludwiga1974sde}, where the random perturbation affects the dynamics continuously. When all such modifications have been applied, the resulting models are called general stochastic hybrid systems (GSHSs) \cite{hu2000towards}. Among these different models, of particular interest for this paper are GHAs with parametric uncertainty and PHAs with additional randomness for both transition probabilities and variable resets. 

\vspace{-.2cm}
When modeling real-world systems using hybrid models, parametric uncertainty arises naturally. Although its cause is multifaceted, two factors are critical. First, probabilistic parameters are needed when the physics controlling the system is known, but some parameters are either not known precisely, are expected to vary because of individual differences, or may change by the end of the system's operational lifetime. Second, system uncertainty may occur when the model is constructed directly from experimental data. Due to imprecise experimental measurements, the values of system parameters may have ranges of variation with some associated likelihood of occurrence. Clearly, the GHAs with parametric uncertainty are suitable models considering these major causes. Note that, in both cases, we assume that the probability distributions of probabilistic system parameters are known. Another interesting and more expressive class of models is PHAs, which extends hybrid automata \cite{henzinger2000theory} with discrete probability distributions. More precisely, for discrete transitions in a model, instead of making a purely nondeterministic choice over the set of currently enabled jumps, a PHA nondeterministically chooses among the set of recently enabled discrete probability distributions, each of which is defined over a set of transitions. Although randomness is defined to only influence the discrete dynamics of the model, PHAs are still very useful and have interesting practical applications \cite{spr2001thesis}. In this paper, we consider a variation of PHAs, where additional randomness for both transition probabilities and resets of some system variables are allowed. In other words, in terms of the randomness for jump probabilities, we mean that the probabilities attached to probabilistic jumps from one mode, instead of having a discrete distribution with predefined constant probabilities, can be expressed by equations involving random variables whose distributions can be either discrete or continuous. This extension is motivated by the fact that some transition probabilities can vary due to factors such as individual and environmental differences in real-world systems. When it comes to the randomness of variable resets, we allow that a system variable can be reset to a value obtained according to a known discrete or continuous distribution, instead of being assigned with a fixed value. For example, with this extension, on a discrete update, variable $t$ can be assigned to any value between $1$ and $2$ with equal probability. 

\vspace{-.2cm}
In this paper, we describe our tool {\it SReach} which supports probabilistic bounded reachability analysis for these two interesting model classes: GHAs with parametric uncertainty and PHAs with additional randomness. It combines the recently proposed $\delta$-complete bounded reachability analysis technique \cite{gaodelta} with statistical testing techniques. Our technique saves the virtues of the Satisfiability Modulo Theories (SMT) based Bounded Model Checking (BMC) for GHAs \cite{cordeiro2012smt, tinelli2012smt}, namely the fully symbolic treatment of hybrid state spaces, while advancing the reasoning power to probabilistic models and requirements. By utilizing the $\delta$-complete analysis method, the full nondeterminism of models can be considered. By adapting statistical tests, {\it SReach} can place arbitrarily small error bounds on the estimated probabilities. Compared to standard simulation-based approaches, our approach supports nondeterministic branching, increases the coverage of simulation, and avoids the zero-crossing problem which is critical for simulation-based methods. Comparing to the existing tools introduced in \cite{zhang2012safety, franzle2008stochastic, david2012statistical, website:prism}, besides offering a sound way to analyze nonlinear dynamics within the SHSs, {\it SReach} also supports probabilistic bounded reachability analysis for hybrid systems with parametric uncertainty. With this modeling formalism, important elements such as probabilistic initial conditions and random variable coefficients can all be expressed by multiple random variables. Furthermore, for PHAs, {\it SReach} considers a more general and useful formalism where general randomness for transition probabilities and variable resets are allowed. We discuss three biological models - a cardiac atrial fibrillation model, a prostate cancer treatment model, and our synthesized Killerred biological model - to show how {\it SReach} can be used to answer several types of questions including model validation, parameter estimation, and sensitivity analysis. To further demonstrate the feasibility of {\it SReach}, we also apply it to additional real-world hybrid systems with parametric uncertainty, e.g. the quadcopter stabilization control.

\vspace{-.2cm}
{\it {\bf Related Work.}} Analysis approaches for GSHSs are often based on Monte-Carlo simulation \cite{blom2004particle}. Considering the difficulty in dealing with this general case, efforts have been mainly placed on different subclasses. For PHAs, Zhang et al. \cite{zhang2012safety} abstracted the original PHA to a probabilistic automaton (PA), and then used established Model Checking methods (e.g. PRISM \cite{website:prism}) for the abstracted model. Hahn et al. also discussed an abstraction-based method where the given PHA was translated into a $n$-player stochastic game using two different abstraction techniques \cite{hahn2011game}. Another method proposed is an SMT-based BMC procedure \cite{franzle2008stochastic}. In \cite{amin2006reachability, abate2007probabilistic, abate2011two, abate2011quantitative}, a similar class of models called discrete-time stochastic hybrid systems (DTSHSs), which is widely used in control theory, was considered. With regard to system analysis, the control problem is to find an optimal control strategy that minimizes the probability of reaching unsafe states. Zuliani et al. also mentioned a simulation-based method for model checking DTSHSs against bounded temporal properties \cite{zuliani2010bayesian}. We refer to this method as Statistical Model Checking (StatMC).  Although StatMC does not belong to the class of exhaustive state-space exploration methods, it usually returns results faster than the exhaustive search with a predefined arbitrarily small error bound on the estimated probability. StatMC was recently integrated into UPPAAL \cite{larsen1997uppaal} in order to handle very general networks of SHAs \cite{david2012statistical}. To analyze reachability problems of SHAs, Fr\"{a}nzle et al. \cite{franzle2011measurability} first over-approximated a given SHA by a PHA, and then exploited the verification procedure introduced in \cite{zhang2012safety} to model check the over-approximating PHA. Plazter introduced another interesting modeling formalism - stochastic hybrid programs (SHPs) in \cite{platzer2011stochastic}. To specify system properties, Platzer proposed a logic called stochastic differential dynamic logic, and then suggested a proof calculus to verify logical properties of SHPs. 

\vspace{-.2cm}
The paper proceeds by first, in Section 2, introducing two modeling formalisms of SHSs under consideration: GHAs with parametric uncertainty, and PHAs with additional randomness. Section 3 explains how {\it SReach} solves the probabilistic bounded reachability problem by encoding stochastic dynamics and combining SMT-based BMC with statistical tests. Case studies and additional experiments are discussed in Section 4. Section 5 concludes the paper.

\vspace{-.1cm}
\section{Stochastic Hybrid Models}
Before discussing the details of the {\it SReach} algorithm, we first define the two types of formalism that {\it SReach} considers. The first class is GHAs with parametric uncertainty. We follow the definition of GHAs in \cite{henzinger2000theory}, and extend it to consider probabilistic parameters in the following way.
\vspace{-.4cm}
\begin{definition}
\label{def:ha_para}
{\rm(Hybrid Automata with Parametric Uncertainty)} A hybrid automaton with probabilistic parameters is a tuple $H_p = \langle (Q, E), \;V, \;RV, \; \mathsf{Init},\; \mathsf{Flow}, \; \mathsf{Inv}, \; \mathsf{Jump}, \; \Sigma \rangle$, where
\vspace{-.4cm}
\begin{itemize}
\item $(Q, E)$ is a finite directed multigraph. The vertices $Q=\{q_1, \cdots,q_m\}$ is a finite set of discrete modes, and edges in $E$ are control switches.
\vspace{-.2cm}
\item $V = \{ v_1, \cdots, v_n \}$ denotes a finite set of real-valued system variables, where $n$ is the dimension of $H_p$. We write $\dot{V}$ for the set $\{\dot{v_1}, \cdots, \dot{v_n}\}$ to represent first derivatives of variables during the continuous change, and write $V'$ for the set $\{v_1', \cdots, v_n'\}$ to denote values of variables at the conclusion of the discrete change.
\vspace{-.2cm}
\item $RV = \{ u_1, \cdots, u_k \}$ is a finite set of random variables, where the distribution of $u_i$ is denoted by $P_i$.
\vspace{-.2cm}
\item $\mathsf{Init}$, $\mathsf{Flow}$, and $\mathsf{Inv}$ are labeling functions over each mode $q \in Q$. The initial condition $Init(q)$ is predicate whose free variables are from $V \cup RV$, the invariant condition $Inv(q)$ is a predicate whose free variables are from $V \cup RV$, and the flow condition $Flow(q)$ is a predicate whose free variables are from $V \cup \dot{V} \cup RV$.
\vspace{-.2cm}
\item $\mathsf{Jump}$ is a transition labeling function that assigns to each transition $e \in E$ a predicate whose free variables are from $V \cup V' \cup RV$.
\vspace{-.2cm}
\item $\Sigma$ is a finite set of events, and an edge labeling function $event: E \to \Sigma$ assigns to each control switch an event. 
\end{itemize}
\end{definition}
\vspace{-.4cm}
{\it SReach} also considers PHAs with additional randomness formally defined as follows.
\vspace{-.4cm}
\begin{definition}
\label{def:pha}
{\rm(Probabilistic Hybrid Automata)} A probabilistic hybrid automaton  (with additional randomness) $H$ is a tuple (${\it Q}$, ${\it \bar{q}}$, ${\it V}$, ${\it \left \langle Post_m \right \rangle_{m \in M}}$, $RV$, ${\it Cmds}$) where
\vspace{-.4cm}
\begin{itemize}
\item ${\it Q} := \{ q_1, \cdots, q_n \}$ is a finite set of control modes.
\vspace{-.2cm}
\item ${\it \bar{q}} \subseteq {\it Q}$ is the initial mode.
\vspace{-.2cm}
\item $V = \{ v_1, \cdots, v_k \}$ denotes a finite set of real-numbered system variables, where $k$ is the dimension of $H$. As mentioned, $\dot{V}$ represents first derivatives of variables, and $V'$ denotes values of variables at the conclusion of the discrete change.
\vspace{-.2cm}
\item ${\it \left \langle Post_q \right \rangle_{q \in Q}}$ indicates continuous-time behaviors on each mode. 
\vspace{-.2cm}
\item $RV$ is a finite set of random variables with known discrete or continuous probability distributions.
\vspace{-.2cm}
\item ${\it Cmds}$ is a finite set of probabilistic guarded commands of the form: 
$g \; \rightarrow \; p_1:u_1 \; + \; \cdots \; + \; p_m:u_m$,
where $g$ is a predicate representing a transition guard with free variables from $V$, $p_i$ is the transition probability for the $i$th probabilistic choice which can be expressed by an equation involving random variable(s) in $RV$ 
and the $p_i$'s satisfy $\sum_{i=1}^m p_i =1$, and $u_i$ is the corresponding transition updating function for the $i$th probabilistic choice, whose free variables are from $V \cup V' \cup RV$.
\end{itemize}
\end{definition}
\vspace{-.4cm}
To illustrate the additional randomness allowed for transition probabilities and variable resets, an example probabilistic guarded command is $x \geq 5 \; \rightarrow \; p_1:(x' = sin(x)) + (1-p_1):(x' = p_x)$, where $x$ is a system variable, $p_1$ has a Uniform distribution $U(0.2, 0.9)$, and $p_x$ has a Bernoulli distribution $B(0.85)$. This means that, the probability to choose the first transition is not a fixed value, but a random one having a Uniform distribution. Also, after taking the second transition, $x$ can be assigned to either $1$ with probability $0.85$, or $0$ with $0.15$. In general, for an individual probabilistic guarded command, the transition probabilities can be expressed by equations of one or more new random variables, as long as values of all transition probabilities are within $[0, 1]$, and their sum is $1$. Currently, all four primary arithmetic operations are supported. Note that, to preserve the Markov property, only unused random variables can be adapted, so that no dependence between the current probabilistic jump and previous transitions will be introduced.

\vspace{-.1cm}
\section{SReach algorithm}
First, {\it SReach} uses a set of random variables to encode all the stochastic information. In detail, when a hybrid automaton with parametric uncertainty is given, {\it SReach} directly declares each probabilistic system parameter as a random variable with a known distribution. While for a PHA, each probabilistic guarded command $g \rightarrow p_1:u_1 + \cdots + p_m:u_m$ is rewritten by introducing a new random variable $rv$ such that $Pr(rv = i) = p_i$. For example, a probabilistic command $x \geq 1 \to 0.7 : (x' = 1)+ 0.3 :( x' = x)$ will be rewritten as two new guarded commands after introducing a new random variable $r$ whose distribution is $(Pr(r=1)=0.7, Pr(r=2)=0.3)$. The first command is $x \geq 1 \wedge r = 1 \to x'=1$. The second is $x \geq 1 \wedge r = 2 \to x'=x$. When additional randomness is involved in assigning probabilities for probabilistic transitions or in resetting system variables, extra random variables are needed. For instance, {\it SReach} can express a probabilistic guarded command as $x \geq 1 \to p_1 : (x' = p_2 )+ (1-p_1) :( x' = x)$, where $p_1$ is a random variable which obeys a Uniform distribution $U(0.6,0.85)$, and $p_2$ is a random variable whose distribution is $N(0,1)$, \ie, normal with mean 0 and standard deviation 1.

\begin{figure}
\centering
\includegraphics[width=\linewidth]{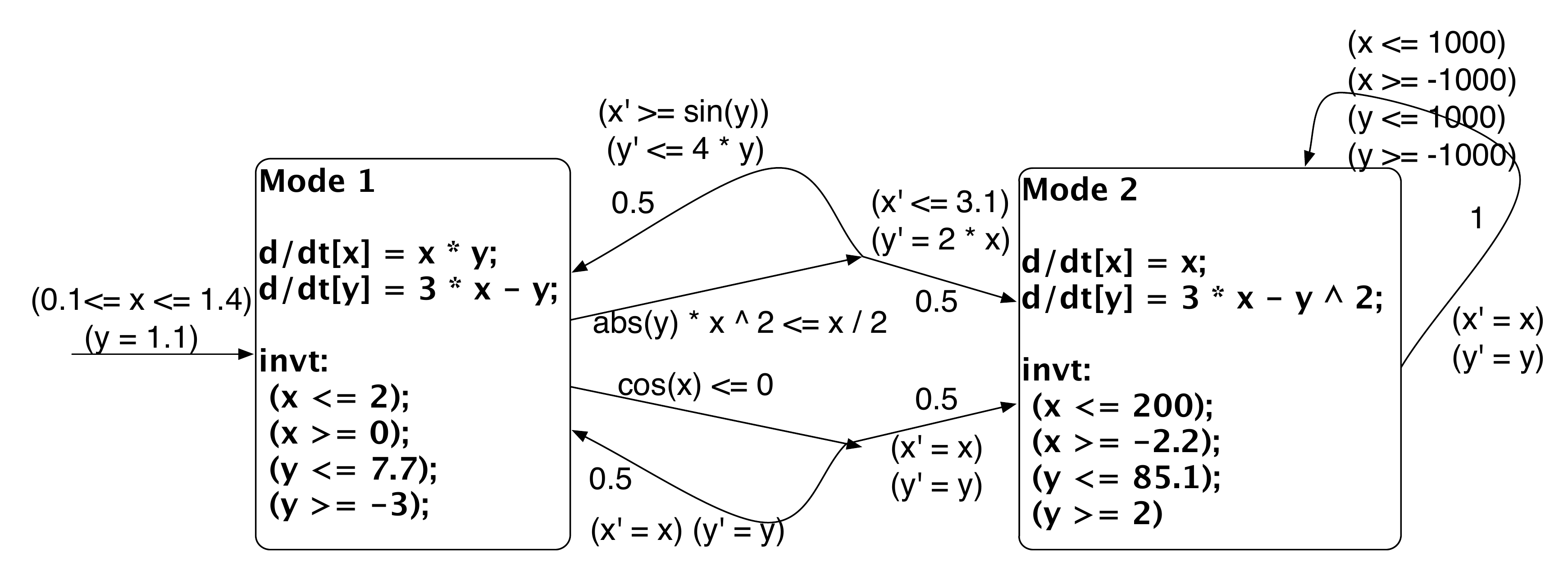}
\caption{An example probabilistic hybrid automaton}
\vspace{-.4cm}
\label{fig:examplepha}
\end{figure}

\vspace{-.2cm}
After encoding all the stochastic elements using random variables, {\it SReach} randomly samples all the random variables according to their probability distributions. For each sampled assignment to these random variables, we obtain a corresponding intermediate hybrid automaton by replacing all the random variables with their assigned values. Then, the $\delta$-complete analyzer {\it dReach} \cite{gaodelta} is utilized to analyze each intermediate hybrid automaton $M_i$, together with the desired precision $\delta$ and unfolding depth $k$. The analyzer returns either unsat or $\delta$-sat for $M_i$ (see Appendix \ref{apndx:dreach} for more details on $\delta$-complete decision procedures). This information is then used by statistical tests to decide whether to stop or to repeat the procedure, and to return the estimated probability. The full procedure is illustrated in Algorithm \ref{fig:sreach}, where $MP$ is a given probabilistic model, and $ST$ indicates which statistical testing method will be used. $Succ$ is used to record the number of $\delta$-sat instances that are returned by {\it dReach}, and $N$ denotes the total number of samples generated so far. These two numbers are then the inputs of {\it SReach}'s statistical testing procedure $ST$. The procedure $ExtractRV$ is used to obtain the full set of involved random variables, $Sim$ is to sample them according to their probability distributions, and $Gen$ is to generate an intermediate hybrid automaton considering the original probabilistic model and a sampled assignment to random variables. Since the full nondeterminism within the intermediate hybrid automata has been considered when handling the bounded reachability problems, the estimated probabilities computed by {\it SReach} are the maximum probabilities. Also, for a probabilistic hybrid automaton, sampling and fixing all the probabilistic transitions in advance results in an over-approximation of the original probabilistic model. Because the result is an over-approximation, safety properties are preserved. To improve the performance of {\it SReach}, each sampled assignment, together with its corresponding {\it dReach} result, has been recorded for avoiding repeated calls to {\it dReach} with the same sampled assignments. This significantly reduces the total calls to {\it dReach} for PHAs (with additional randomness for transition probabilities), as the size of the sample space involving random variables describing probabilistic jumps is comparatively small. For the example PHA, as shown in Figure \ref{fig:examplepha}, with this improvement, the total checking time for a reachability problem has been decreased from $11291.31$s for $658$ samples ($17.16$s per sample) to $3295.82$s ($5.01$s per sample). To further improve the performance, a parallel version of {\it SReach} has been implemented using OpenMP, where multiple samples and corresponding hybrid automata are generated, and passed to {\it dReach} simultaneously. Using this parallel {\it SReach} on a 4-core machine, the running time for the example PHA has been further decreased to $2119.55s$ for $660$ samples ($3.33$s per sample). 
\vspace{-.2cm}

\begin{algorithm}
  \centering
  \caption{SReach}
  \label{fig:sreach}
  \begin{algorithmic}[1]
    \Function{SReach}{$M\!P$, $ST$, $\delta$, $k$}
        \State $Succ \gets 0$	
        \State $N \gets 0$
        \State $RV \gets \mathrm{ExtractRV}(MP)$	
        \Repeat
            \State $S_i \gets \mathrm{Sim}(RV)$		
            \State $M_i \gets \mathrm{Gen}(M\!P, S_i)$	
            \State $Res \gets \mathrm{dReach}(M_i, \delta, k)$	
            \If{$Res$ = $\delta$-sat}
		\State $Succ \gets Succ + 1$
	    
	  \EndIf
	\State $N \gets N + 1$
        \Until{$ST.done(Succ, N)$}\\
	\quad\hspace{0.5ex} \Return $ST.output$
   \EndFunction
  \end{algorithmic}
\end{algorithm}

Currently, {\it SReach} supports a number of hypothesis testing and statistical estimation techniques including: {Lai's test} \cite{lai1988nearly}, {Bayes factor test} \cite{kass1995bayes}, {Bayes factor test with indifference region} \cite{younes2005verification}, {Sequential probability ratio test (SPRT)}\cite{wald1945sequential}, {Chernoff-Hoeffding bound} \cite{hoeffding1963probability}, {Bayesian Interval Estimation with Beta prior}\cite{zuliani2010bayesian}, and {Direct Sampling}. All methods produce answers which are correct up to a precision that can be set arbitrarily by the user. See Appendix \ref{apndx:stat} for more details about how these tests can guarantee an arbitrary small error bound between the estimated probability and the real one. With these hypothesis testing methods, {\it SReach} can answer qualitative questions, such as ``Does the model satisfy a given reachability property in $k$ steps with probability greater than a certain threshold?'' With the above statistical estimation techniques, {\it SReach} can offer answers to quantitative problems. For instance, ``What is the probability that the model satisfies a given reachability property in $k$ steps?''  {\it SReach} can also handle additional types of interesting problems by encoding them as bounded reachability problems. The {\bf model validation} problem with prior knowledge can be encoded as a bounded reachability question. After expressing prior knowledge about the given model as reachability properties, is there any number of steps $k$ in which the model satisfies a given property? If none exists, the model is incorrect regarding the given prior knowledge. If, for each property, a witness is returned, we can conclude that the model is correct with regard to the prior knowledge. The {\bf parameter estimation} problem can also be encoded as a $k$-step reachability problem. Does there exist a parameter combination for which the model reaches the given goal region in $k$ steps? If so, this parameter combination is potentially a good estimation for the system parameters. The goal here is to find a combination with which all the given goal regions can be reached in a bounded number of steps. Moreover, {\bf sensitivity analysis} can be conducted by a set of bounded reachability queries as well: Are the results of reachability analysis the same for different possible values of a certain system parameter? If so, the model is insensitive to this parameter with regard to the given prior knowledge.

\section{Experiments}
Our method is implemented in the open-source tool {\it SReach} (\url{https://github.com/dreal/SReach}). Both a sequential version and a parallel one have been implemented. See Appendix \ref{apndx:usage} for information on using {\it SReach}. All models for the following case studies and additional benchmarks can be found on the tool website. All experiments were conducted on a server with 2* AMD Opteron(tm) Processor 6172 (24 cores) and 32GB RAM, running on Ubuntu 14.04.1 LTS. 12 cores were used. In our experiments we used $0.001$ as the precision for the $\delta$-decision problem, and Bayesian sequential estimation with $0.01$ as the estimation error bound, coverage probability $0.99$, and a uniform prior ($\alpha = \beta = 1$).

\vspace{-.1cm}
{\bf\noindent Atrial Fibrillation.} The minimum resistor model (MRM) reproduces experimentally measured characteristics of human ventricular cell dynamics \cite{bueno2008minimal}. The MRM reduces the complexity of existing models by representing channel gates of different ions with one fast channel, and two slow gates. However, due to this reduction, for most model parameters, it becomes impossible to obtain their values through measurements. After adding parametric uncertainty into the original hybrid model, we show that {\it SReach} can be adapted to estimate parameters for this stochastic model, i.e., identifying appropriate ranges and distributions for model parameters. To illustrate the way in which {\it SReach} is used to conduct parameter estimation, we chose two system parameters - $E\!P\!I\_\!T\!O1$ and $E\!P\!I\_\!T\!O2$, and varied their distributions to see which ones allow the model to present the desired pattern. The model has 4 modes. In the experiments, we chose $3$ as the unfolding depth. For each sample generated, {\it SReach} analyzed systems with $62$ variables, and $24$ ODEs. As in Table \ref{table:cardiac}, when $E\!P\!I\_\!T\!O1$ is either close to $400$, or between $0.0061$ and $0.007$, and $E\!P\!I\_\!T\!O2$ is close to $6$, the model can satisfy the given bounded reachability property with a probability very close to $1$. 
\vspace{-.2cm}
\begin{table}[h!]
\captionsetup{font=scriptsize}
\centering
    \begin{tabular}{|p{1.3cm} < {\centering}|p{1cm} < {\centering}|p{0.7cm} < {\centering}|p{0.7cm} < {\centering}|p{0.7cm} < {\centering}|p{0.7cm} < {\centering}|p{0.7cm} < {\centering}|}
    \hline
    \small{EPI\_\!T\!O1}            &\small{EPI\_\!T\!O2}         &\small{\#S\_S} & \small{\#T\_S} & \small{Est\_P} & \small{A\_T(s)} & \small{T\_T(s)} \\ \hline
    U(6.1e-3, 7e-3)    & 6              & 240       & 240      & 0.996     & 0.270   & 64.80     \\ \hline
    U(5.5e-3, 5.9e-3)   & 6              & 0         & 240      & 0.004     & 0.042  & 10.08       \\ \hline
    400               & U(0.131, 6)    & 240      & 240      & 0.996     & 0.231  & 55.36      \\ \hline
    400               & U(0.1, 0.129)    & 0         & 240      & 0.004     & 0.038   & 9.15     \\ \hline
    N(400, 1e-4)      & N(6, 1e-4)     & 240       & 240      & 0.996     & 0.091  & 21.87      \\ \hline
    N(5.5e-3, 10e-6) & N(0.11, 10e-5) & 0         & 240      & 0.004     & 0.037  & 8.90      \\ \hline
    \end{tabular}
    \caption {Results for the atrial fibrillation model. \#RVs = number of random variables in the model, \#S\_S = number of $\delta$-sat samples, 
\#T\_S = total number of samples, Est\_P = estimated maximum posterior probability,  A\_T(s) = average 
CPU time of each sample in seconds, and T\_T(s) = total CPU time for all samples in seconds.}
\vspace{-.5cm}
    \label{table:cardiac}
\end{table}

{\bf\noindent Prostate cancer treatment.}
This model is a nonlinear hybrid automaton with parametric uncertainty. We modified the model of the intermittent androgen suppression (IAS) therapy in \cite{tanaka2010mathematical} by adding parametric uncertainty. The IAS therapy switches between  treatment-on, and treatment-off with respect to the serum level thresholds of prostate-specific antigen (PSA), namely $r_0$ and $r_1$. As suggested by the clinical trials \cite{bruchovsky2006final}, an effective IAS therapy highly depends on the individual patient. Thus, we modified the model by taking parametric variation caused by personalized differences into account. In detail, according to clinical data from hundreds of patients \cite{bruchovsky2007locally}, we replaced six system parameters with random variables having appropriate (continuous) distributions, including $\alpha_x$ (the proliferation rate of androgen-dependent (AD) cells), $\alpha_y$ (the proliferation rate of androgen-independent (AI) cells), $\beta_x$ (the apoptosis rate of AD cells), $\beta_y$ (the apoptosis rate of AI cells), $m_1$ (the mutation rate from AD to AI cells), and $z_0$ (the normal androgen level). To describe the variations due to individual differences, we assigned $\alpha_x$ to be $U(0.0193, 0.0214)$, $\alpha_y$ to be $U(0.0230, 0.0254)$, $\beta_x$ to be $U(0.0072, 0.0079)$, $\beta_y$ to be $U(0.0160, 0.0176)$, $m_1$ to be $U(0.0000475, 0.0000525) $, and $z_0$ to be $N(30.0, 0.001)$. We used {\it SReach} to estimate the probabilities of preventing the relapse of prostate cancer with three distinct pairs of treatment thresholds (\ie, combinations of $r_0$ and $r_1$).  In the experiments, we chose $k=2$ as the unfolding depth. For each sample generated, {\it SReach} analyzed systems with $41$ variables, and $10$ ODEs. As shown in Table \ref{table:prostate}, the model with thresholds $r_0 = 10$ and $r_1 = 15$ has a maximum posterior probability that approaches 1, indicating that these thresholds may be considered for the general treatment. 
\begin{table}[th!]
\captionsetup{font=scriptsize}
\centering
    \begin{tabular}{|p{1.1cm} < {\centering}|p{0.65cm} < {\centering}|p{0.65cm} < {\centering}|p{0.65cm} < {\centering}|c|c|c|c|c|}
    \hline
    $(r_0,r_1)$ & Est\_P & \#S\_S & \#T\_S & Avg\_T(s) & Tot\_T(s) \\ \hline
    (5, 10) & 0.496   & 8226      & 16584    & 0.596   & 9892     \\ \hline
    (7, 11) & 0.994  & 335   & 336   & 54.307 & 18247     \\ \hline
    (10, 15) & 0.996  & 240    & 240    & 506.5   & 121560   \\ \hline
    \end{tabular}
    \caption{Results for the prostate cancer treatment model. \#S\_S = number of $\delta$-sat samples, 
\#T\_S = total number of samples, $r_0$ = lower threshold of the serum PSA level, $r_1$ = upper threshold, 
Est\_P = estimated maximum posterior probability,  Avg\_T(s) = average CPU time of each sample in seconds, and Tot\_T(s) = total CPU time for all samples in seconds.}
    \label{table:prostate}
\end{table}
\vspace{-.65cm}

{\bf Synthesized Killerred Model.} Due to the widespread misuse and overuse of antibiotics, drug resistant bacteria now pose significant risks to health, agriculture and the environment. An alternative to conventional antibiotics is phage-based therapy. Our approach to antibiotic resistance is to engineer a temperate phage, Lambda ($\lambda$), with light-activated production of superoxide (SOX). We incorporated the Killerred protein which has been shown to be phototoxic, and can provide another level of controlled bacteria killing \cite{natasa2014killerred}. A probabilistic hybrid automaton for this synthesized Killerred model, as shown in Figure \ref{fig:killerred} in Appendix \ref{apndx:model}, has been constructed. Considering individual differences of bacterial cells and distinct experimental environments, additional randomness on transition probabilities were considered. {\it SReach} was first used to validate this model by estimating the probabilities of killing bacterial cells with different values for $k$, as shown in Table \ref{table:kr01}. We noticed that the probabilities of paths going through mode $6$ to mode $11$ in Figure \ref{fig:killerred} are close to $0$. To exclude the effect from sampling of rare events, we increased the probability of entering mode $6$. After this modification, the corresponding probabilities estimated by {\it SReach} still approach $0$. We conclude that it is impossible for this model to enter mode $6$. {\it SReach} was also used to (a) find out the relation between the time to turn on the light after adding the molecular biology reagent IPTG and the total time to kill bacterial cells (see Table \ref{table:kr02} in Appendix \ref{apndx:exp}), (b) figure out that the lower bound for the duration of exposure to light is $3$ (see Table \ref{table:kr03} in Appendix \ref{apndx:exp}), (c) find that the time to remove IPTG is not sensitive considering whether bacterial cells will be killed, and (d) estimate that the upper bound of $SOX_{thres}$ (the necessary concentration of SOX to kill bacterial cells) is $0.6667$. All these findings have been reported to biologists for further checking.

\vspace{-.2cm}
\begin{table}[th!]
\captionsetup{font=scriptsize}
\centering
    \begin{tabular}{|p{1.1cm} < {\centering}|p{0.65cm} < {\centering}|p{0.65cm} < {\centering}|p{0.65cm} < {\centering}|c|c|c|c|c|}
    \hline
    $k$ & Est\_P & \#S\_S & \#T\_S & Avg\_T(s) & Tot\_T(s) \\ \hline
    5 &  0.544  & 8951     &  16452   & 0.074   & 1219.38     \\ \hline
    6 & 0.247  & 3045   & 12336   & 0.969 & 11957.12     \\ \hline
    7 & 0.096  & 559    & 5808    & 5.470   & 31770.36   \\ \hline
    8 & 0.004  & 0      & 240    & 0.004  & 0.88     \\ \hline
    9 & 0.004  & 0   & 240   & 0.012 & 2.97     \\ \hline
    10 & 0.004  & 0    & 240    & 0.013   & 3.18   \\ \hline
    \end{tabular}
    \caption{Results for the killerred model. \#S\_S = number of $\delta$-sat samples, 
\#T\_S = total number of samples, $r_0$ = lower threshold of the serum PSA level, $r_1$ = upper threshold, 
Est\_P = estimated maximum posterior probability,  Avg\_T(s) = average CPU time of each sample in seconds, and Tot\_T(s) = total CPU time for all samples in seconds.}
    \label{table:kr01}
\end{table}
\vspace{-.3cm}

{\bf Additional benchmarks.} To further demonstrate the feasibility of {\it SReach}, we also applied it to additional benchmarks including hybrid systems with parametric uncertainty, PHAs, and PHAs with additional randomness. Appendix \ref{apndx:exp} shows the results of these experiments. Moreover, the detailed description of some of the additional benchmarks and above case studies are presented in Appendix \ref{apndx:model}.

\section{Conclusions and future work}
We have presented a probabilistic bounded reachability analysis tool that combines $\delta$-decision 
procedures and statistical tests. It supports reachability analysis for hybrid systems with parametric 
uncertainty and probabilistic hybrid automata with additional randomness. It takes the full nondeterminism of models into account, and ensures the estimation accuracy by placing arbitrary small error bounds. This tool has been used for the probabilistic bounded reachability analysis of three representative examples - a prostate 
cancer treatment model, a cardiac model, and a synthesized Killerred model - which are currently out of the reach of other formal tools. In the near future, we plan to extend support for more general stochastic hybrid models 
that include probabilistic jumps with continuous distributions, and stochastic differential equations.

\vspace{-.4cm}
\bibliography{ref}{}
\bibliographystyle{abbrv}

\newpage
\appendix

\section{Statistical tests}\label{apndx:stat}
In this section we briefly describe the statistical techniques implemented in {\it SReach}.
To deal with qualitative questions, {\it SReach} supports the following hypothesis testing methods.

\textit{Lai's test} \cite{lai1988nearly}.
As a simple class of sequential tests, it tests the one-sided composite hypotheses $H_0: \; \theta \leq \theta_0$ versus $H_1:\; \theta \geq \theta_1$ for the natural parameter $\theta$ of an exponential family of distributions under the $0-1$ loss and cost $c$ per observation. \cite{lai1988nearly} shows that these tests have nearly optimal frequentist properties and also provide approximate Bayes solutions with respect to a large class of priors. 

\textit{Bayes factor test} \cite{kass1995bayes}.
The use of Bayes factors is a Bayesian alternative to classical hypothesis testing. It is based on the Bayes theorem. Hypothesis testing with Bayes factors is more robust than frequentist hypothesis testing, as the Bayesian form avoids model selection bias, evaluates evidence in favor of the null hypothesis, includes model uncertainty, and allows non-nested models to be compared. Also, frequentist significance tests become biased in favor of rejecting the null hypothesis with sufficiently large sample size. 

\textit{Bayes factor test with indifference region}. 
A hypothesis test has ideal performance if the probability of the Type-I error (respectively, Type-II error) is exactly $\alpha$ (respectively, $\beta$). However, these requirements make it impossible to ensure a low probability for both types of errors simultaneously (see \cite{younes2005verification} for details). A solution is to use an indifference region. The indifference region indicates the distance between two hypotheses, which is set to separate the two hypotheses.

\textit{Sequential probability ratio test (SPRT)} \cite{wald1945sequential}. 
The SPRT considers a simple hypothesis $H_0:\;\theta = \theta_0$ against a simple alternative $H_1:\;\theta = \theta_1$. With the critical region $\Lambda_n$ and two thresholds $A$, and $B$, SPRT decides that $H_0$ is true and stops when $\Lambda_n < A$. It decides that $H_1$ is true and terminates if $\Lambda_n > B$. If $A\; < \Lambda_n < B$, it will collect another observation to obtain a new critical region $\Lambda_{n+1}$. The SPRT is optimal, among all sequential tests, in the sense that it minimizes the average sample size.

To offer quantitative answers, {\it SReach} also supports estimation procedures as below.

\textit{Chernoff-Hoeffding bound} \cite{hoeffding1963probability}. To estimate the mean $p$ of a (bounded) 
random variable, given a precision $\delta'$ and coverage probability $\alpha$, the Chernoff-Hoeffding bound 
computes a value $p'$ such that $|p' \; - \; p| \le \delta'$ with probability at least $\alpha$.

\textit{Bayesian Interval Estimation with Beta prior} \cite{zuliani2010bayesian}. This method estimates $p$, the unknown probability that a random sampled model satisfies a specified reachability property. 
The estimate will be in the form of a confidence interval, containing $p$ with an arbitrary high probability.  \cite{zuliani2010bayesian} assumes that the unknown $p$ is given by a random variable, whose density is called the prior density, and focuses on Beta priors. 

\textit{Direct sampling}. Given $N$ as the number of samples to be sampled, the direct sampling method estimates the mean of $p$ of a (bounded) random variable. According to the central limit thoerem \cite{durrett2010probability}, the error $\epsilon$ with a confidence $c$ between the real probability $p$ and the estimated $\hat{p}$ is bounded: \\
$\epsilon  =  \phi^{-1}\left ( \frac{c + 1}{2} \right ) \sqrt{\frac{p(1-p)}{N}}$\\
where $\phi(x) = \frac{1}{\sqrt{2\pi}} \int_{-x}^{x} e^{-t^2 / 2}dt$. That is, as $N$ goes to $\infty$, the estimated probability approaches to the real one.

\section{$\delta$-Decisions for Hybrid Models}\label{apndx:dreach}
The reachability problems of hybrid automata can be encoded using a first-order language $\lrf$ over the reals,
which allows the use of a wide range of real functions including nonlinear ODEs.
Then, $\delta$-complete decision procedures are used to find solutions to these formulas to synthesize
parameters.
\begin{definition}[$\lrf$-Formulas]
Let $\mathcal{F}$ be a collection of computable real functions. We define:
\begin{align*}
t& := x \; | \; f(t(\vec x)), \mbox{ where }f\in \mathcal{F} \mbox{ (constants are 0-ary functions)};\\
\varphi& := t(\vec x)> 0 \; | \; t(\vec x)\geq 0 \; | \; \varphi\wedge\varphi
\; | \; \varphi\vee\varphi \; | \; \exists x_i\varphi \; |\; \forall x_i\varphi.
\end{align*}
\end{definition}
By computable real function we mean Type 2 computable, which informally requires that a (real)
function can be algorithmically evaluated with arbitrary accuracy. Since in general
$\lrf$ formulas are undecidable, the decision problem needs to be relaxed. In particular, for
any $\lrf$ formula $\phi$ and any rational $\delta >0$ one can obtain a $\delta$-weakening
formula $\phi^\delta$ from $\phi$ by substituting the atoms $t > 0$ with $t > -\delta$ (and
similarly for $t \geq 0$). Obviously, $\phi$ implies $\phi^\delta$, but not the {\em vice versa}.
Now, the $\delta$-decision problem is deciding correctly whether:
\begin{itemize}
\item $\phi$ is false ($\mathsf{unsat}$);
\item $\phi^\delta$ is true ($\delta$-$\mathsf{sat}$).
\end{itemize}
If both cases are true, then either decision is correct. More details on algorithms ($\delta$-{\em complete} decision procedures) for solving $\delta$-decision problems for $\lrf$ and for ODEs can be found in \cite{gao2013satisfiability, gao2013dreal, gaodelta}.

Now we state the encoding for hybrid models. Hybrid automata generalize finite-state
automata by permitting continuous time flow in each discrete mode.
Also, in each mode an {\em invariant} must be satisfied by the flow, and mode switches are controlled
by {\em jump} conditions.
\begin{definition}[$\lrf$-Representations of Hybrid Automata]\label{lrf-definition}
A hybrid\\ automaton in $\lrf$-representation is a tuple
\begin{multline*}
H = \langle X, Q, \{{\flow}_q(\vec x, \vec y, t): q\in Q\},\{\inv_q(\vec x): q\in Q\},\\
\{\jump_{q\rightarrow q'}(\vec x, \vec y): q,q'\in Q\},\{\init_q(\vec x): q\in Q\}\rangle
\end{multline*}
where $X\subseteq \mathbb{R}^n$ for some $n\in \mathbb{N}$, $Q=\{q_1,...,q_m\}$ is a finite set of modes, and the other components are finite sets of quantifier-free $\lrf$-formulas.
\end{definition}

We now show the encoding of bounded reachability, which is used for encoding the parameter synthesis
problem. We want to decide whether a given
hybrid system reaches a particular region of its state space after following a (bounded) number
of discrete transitions, \ie, jumps. First, we need to define auxiliary formulas used
for ensuring that a particular mode is picked at a certain step.
\begin{definition}
Let $Q = \{q_1,...,q_m\}$ be a set of modes. For any $q\in Q$, and $i\in\mathbb{N}$, use $b_{q}^i$ to represent a Boolean variable. We now define
$$\enforce_Q(q,i) = b^i_{q} \wedge \bigwedge_{p\in Q\setminus\{q\}}\neg b^{i}_{p}$$
$$\enforce_Q(q, q',i) = b^{i}_{q}\wedge \neg b^{i+1}_{q'} \wedge \bigwedge_{p\in Q\setminus\{q\}} \neg b^i_{p} \wedge \bigwedge_{p'\in Q\setminus\{q'\}} \neg b^{i+1}_{p'}$$
We omit the subscript $Q$ when the context is clear.\end{definition}
We can now define the following formula that checks whether a {\em goal} region of the automaton
state space is reachable after exactly $k$ discrete transitions. We first state
the simpler case of a hybrid system without invariants.
\begin{definition}[$k$-Step Reachability, Invariant-Free Case]
Suppose $H$ is an invariant-free hybrid automaton, $U$ a subset of its state space represented by $\goal$,
and $M>0$. \\The formula $\reach_{H,U}(k,M)$ is defined as:
\begin{eqnarray*}
& &\exists^X \vec x_{0} \exists^X\vec x_{0}^t\cdots \exists^X \vec x_{k}\exists^X\vec x_{k}^t\exists^{[0,M]}t_0\cdots \exists^{[0,M]}t_k.\\
& &\bigvee_{q\in Q} \Big(\init_{q}(\vec x_{0})\wedge \flow_{q}(\vec x_{0}, \vec x_{0}^t, t_0)\wedge \enforce(q,0)\Big)\\
\wedge & & \bigwedge_{i=0}^{k-1}\bigg( \bigvee_{q, q'\in Q} \Big(\jump_{q\rightarrow q'}(\vec x_{i}^t, \vec x_{i+1})\wedge \enforce(q,q',i)\\
& & \hspace{1.1cm}\wedge\flow_{q'}(\vec x_{i+1}, \vec x_{i+1}^t, t_{i+1}) \wedge \enforce(q',i+1)\Big)\bigg)\\
\wedge & &\bigvee_{q\in Q} (\goal_q(\vec x_{k}^t)\wedge \enforce(q,k))
\end{eqnarray*}
where $\exists^X x$ is a shorthand for $\exists x\in X$.
\end{definition}
Intuitively, the trajectories start with some initial state satisfying $\init_q(\vec x_{0})$ for some $q$.
Then, in each step the trajectory follows $\flow_q(\vec x_{i}, \vec x_{i}^t, t)$ and makes a continuous flow from $\vec x_i$ to $\vec x_i^t$ after time $t$. When the automaton makes a $\jump$ from mode $q'$ to $q$, it resets variables following $\jump_{q'\rightarrow q}(\vec x_{k}^t, \vec x_{k+1})$. The auxiliary $\enforce$ formulas ensure that picking $\jump_{q\rightarrow q'}$ in the $i$-the step enforces picking $\flow_q'$ in the $(i+1)$-th step.
When the invariants are not trivial, we need to ensure that for all the time points along a continuous flow, the invariant condition holds. We need to universally quantify over time, and the encoding is as follows:
\begin{definition}[$k$-Step Reachability, Nontrivial Invariant]\label{br2}
Suppose $H$ contains invariants, and $U$ is a subset of the state space represented by $\goal$. The $\lrf$-formula $\reach_{H,U}(k,M)$ is defined as:
\begin{eqnarray*}
& &\exists^X \vec x_{0} \exists^X\vec x_{0}^t\cdots \exists^X \vec x_{k}\exists^X\vec x_{k}^t \exists^{[0,M]}t_0\cdots \exists^{[0,M]}t_k.\\
& &\bigvee_{q\in Q} \Big(\init_{q}(\vec x_{0})\wedge \flow_{q}(\vec x_{0}, \vec x_{0}^t, t_0)\wedge \enforce(q,0)\\
& &\hspace{1.1cm} \wedge \forall^{[0,t_0]}t\forall^X\vec x\;(\flow_{q}(\vec x_{0}, \vec x, t)\rightarrow \inv_{q}(\vec x))\Big) \\
\wedge & &\bigwedge_{i=0}^{k-1}\bigg( \bigvee_{q, q'\in Q} \Big(\jump_{q\rightarrow q'}(\vec
x_{i}^t, \vec x_{i+1})\wedge \flow_{q'}(\vec x_{i+1}, \vec x_{i+1}^t, t_{i+1})\\
& &\hspace{1.1cm} \wedge \enforce(q,q',i) \wedge\enforce(q',i+1)\\
& &\hspace{1.1cm} \wedge \forall^{[0,t_{i+1}]}t\forall^X\vec x\;(\flow_{q'}(\vec x_{i+1}, \vec x,
t)\rightarrow \inv_{q'}(\vec x)) )\Big)\bigg)\\
\wedge & &\bigvee_{q\in Q} (\goal_q(\vec x_{k}^t)\wedge \enforce(q,k)).
\end{eqnarray*}
\end{definition}
The extra universal quantifier for each continuous flow expresses the requirement that for all the time points between the initial and ending time point ($t\in[0,t_i+1]$) in a flow, the continuous variables $\vec x$ must take values that satisfy the invariant conditions $\inv_q(\vec x)$.

\section{The {\it SReach} tool}\label{apndx:usage}
\subsection{Input format}
The inputs to our {\it SReach} tool are descriptions of (probabilistic) hybrid automata with random variables (representing the probabilistic system parameters, and probabilistic jumps), and the reachability property to be checked. Following roughly the same format as the above definition of (probabilistic) hybrid automata, and adding the declarations of random variables, the description of an automaton is as follows.

{\bf Preprocessor.} We can use the C language syntax to define constants and macros. 

{\bf Variable declaration.} For a random variable, the declaration specifies its distribution and name. Variables which are not random variables are required to be declared within bounds. 

{\bf (Probabilistic) Hybrid automaton.} A (probabilistic) hybrid automaton is represented by a set of modes. Within each mode declaration, we can specify statements for the mode invariant(s), flow function(s), and (probabilistic) jump condition(s). For a mode invariant, we can give any logic formula of the variables. A flow function is expressed by an ODE.  As for a nonprobabilistic jump condition, it is written as 
\vspace{-0.4cm}
\begin{verbatim} 
<logic_formula1>  ==> 
            @<target_mode>  <logic_formula2>,
\end{verbatim}
\vspace{-0.4cm}
where the first logic formula is given as the guard of the jump, and the second one specifies the reset condition after the jump. While for a probabilistic jump condition, we need an extra constraint to express the stochastic choice, which is of the following form
\vspace{-0.4cm}
\begin{verbatim}
(and <logic_formula1> <stochastic choice>)  ==> 
            @<target_mode>  <logic_formula2>,
\end{verbatim}
\vspace{-0.4cm}
where the stochastic choice is a formula indicating which probabilistic transition will be chosen for this jump.

{\bf Initial conditions and Goals.} Following the declaration of modes, we can declare one initial mode with corresponding conditions, and the reachability properties in the end.

\noindent\textit{Example 1}. The following is an example input file for a hybrid automaton with parametric uncertainty. Currently, users can specify random variables (representing certain system parameters) with Bernoulli distribution (B), Uniform distribution (U), Gaussian distribution (N), Exponential distribution (E), and general Discrete distribution with given possible values and corresponding probabilities (DD). 
\lstset{basicstyle=\ttfamily\small, numbers=left, breaklines=true }
\begin{lstlisting}
#define pi 3.1416
N(1,0.1) mu1;
U(10,15) thro;
E(0.49) theta1;
B(0.75) xinit;
DD(0:0.7, 1:0.3) mu2;
[0,5] x;
[0,3] time;
{ mode 1; 
  invt:
         (x<=1.5);
         (x>=0);
  flow:
	 d/dt[x]=thro*(1/(theta1*sqrt(2*pi)))
	          *exp(0-((x-mu1+mu2)^2)/(2*theta1^2));
  jump:
        (x>=(thre1+5))==>@2(x'=x);
}
init:
@1	(x=xinit);
goal:
@4	(x>=50);
\end{lstlisting}

\noindent\textit{Example 2}.  This example demonstrates the format of the input file for a probabilistic hybrid automaton with additional randomness for transition probabilities. Note that, unlike the notations of declarations of random variables representing system parameters and probabilistic transitions, declarations of random variables used to express the addtional randomness for jump probabilities start with a prefix $j$.
\lstset{basicstyle=\ttfamily\small, numbers=left, breaklines=true }
\begin{lstlisting}
jU(0.7, 0.9) pjumprv;
DD(1:pjumprv, 2:(1 - pjumprv)) pjump1;
DD(1:0.3, 2:0.7) pjump2;
[-1000, 1000] x;
[-1000, 1000] y;
[0, 3] time;

{ mode 1;

  invt:
        (x <= 2);
        (x >= 0);
	(y <= 7.7);
	(y >= -3);
  flow:
        d/dt[x] = x * y;
        d/dt[y] = 3 * x - y;
  jump:
        (and (abs(y) * x ^ 2 <= x / 2) (pjump1 = 1)) ==> @1 (and (x' >= sin(y)) (y' <= 4 * y));
	(and (abs(y) * x ^ 2 <= x / 2) (pjump1 = 2)) ==> @2 (and (x' <= 3.1) (y' = 2 * x));
	(and (cos(x) <= 0) (pjump2 = 1)) ==> @2 (and (x' = x) (y' = y));
	(and (cos(x) <= 0) (pjump2 = 2)) ==> @1 (and (x' = x) (y' = y));
}

{
  mode 2;
  invt:
        (x <= 200);
        (x >= -2.2);
	(y <= 85.1);
	(y >= 2);
  flow:
        d/dt[x] = x;
        d/dt[y] = 3 * x - y ^ 2;
  jump:
        (and (x <= 1000) (x >= -1000) (y <= 1000) (y >= -1000)) ==> @2 (and (x' = x) (y' = y));
}
init:
@1	(and (x >= 0.1) (x <= 1.4) (y = 1.1));

goal:
@2	(and (x >= -10) (y >= -10));
\end{lstlisting}

\subsection{Command line}
{\it SReach} offers two choices. It can be run sequentially by typing
\vspace{-0.4cm}
\begin{verbatim} 
sreach_sq <statistical_testing_option> <filename> 
		<dReach> <k> <delta>,
\end{verbatim} 
\vspace{-0.4cm}
or in parallel by 
\vspace{-0.4cm}
\begin{verbatim} 
sreach_para <statistical_testing_option> <filename> 
		<dReach> <k> <delta>,
\end{verbatim} 
\vspace{-0.4cm}
where:
\begin{itemize}
\item \verb+statistical_testing_option+ is a text file containing a sequence of test specifications. We will introduce the usages of statistical testing options in the following part;
\item \verb+filename+ is a .pdrh file describing the model of a hybrid system with probabilistic system parameters. It is of the input format described in last sub-section;
\item \verb+dReach+ is a tool for bounded reachability analysis of hybrid systems based on dReal;
\item \verb+k+ is the number of steps of the model that the tool will explore; and
\item \verb+delta+ is the precision for the $\delta$-decision problem.
\end{itemize}

\subsection{Statistical testing options}

{\it SReach} can be used with different statistical testing methods through the following specifications.

\textit {Lai's test}: \verb+Lai <theta> <cost_per_sample>+, where \verb+theta+ indicates the probability threshold.
 
\textit {Bayes factor test}: \verb+BFT <theta> <T> <alpha> <beta>+,
where \verb+theta+ is a probability threshold satisfying \verb+0 < theta < 1+, \verb+T+ is a ratio threshold satisfying \verb+T > 1+, and \verb+alpha+, and \verb+beta+ are beta prior parameters.

\textit {BFT with indifference region}: \\ \verb+BFTI <theta> <T> <alpha> <beta> <delta>+,
where, besides the parameters used in the above Bayes factor test, \verb+delta+ is given to create the indifference region - [$p_0$, $p_1$], where $p_0$ = \verb+theta+ - \verb+delta+ and $p_1$ = \verb+theta+  + \verb+delta+.  Now, it tests $H_0 :\; p \ge p_0$ against $H_1:\; p \le p_1$ .

\textit {Sequential probability ratio test (SPRT)}: \\ \verb+SPRT <theta> <T> <delta>+.

\textit {Chernoff-Hoeffding bound}:\\ \verb+CHB <delta1> <coverage_probability>+,
where \verb+delta1+ is the given precision, and \verb+coverage_probability+ indicates the confidence.

\textit {Bayesian Interval Estimation with Beta prior}: \\ \verb+BEST <delta1> <coverage_probability> <alpha> <beta>+.

\textit {Direct/Na\"{i}ve Sampling}: \verb+NSAM <num_of_samples>+.

\vspace{-2.2cm}
\section{Model description}\label{apndx:model}

\textbf{\textit{Atrial Fibrillation.}} The model has four discrete control locations, four state variables, and nonlinear ODEs. A typical set of ODEs in the model is:
\begin{eqnarray*}
\frac{du}{dt} &=& e + (u-\theta_v)(u_u-u ) v g_{fi} + wsg_{si}-g_{so}(u)\\
\frac{ds}{dt} &=& \displaystyle\frac{g_{s2}}{(1+\exp(-2k(u-us)))} -  g_{s2}s\\
\frac{dv}{dt} &=& -g_v^+\cdot v \hspace{1cm} \frac{dw}{dt} = -g_w^+\cdot w
\end{eqnarray*}
The exponential term on the right-hand side of the ODE is the sigmoid function, which often appears in modelling biological switches.

\textbf{\textit{Electronic Oscillator.}} The 3dOsc model represents an electronic oscillator model that contains nonlinear ODEs such as the following.
\begin{eqnarray*}
\frac{dx}{dt} &=& - ax \cdot sin(\omega_1 \cdot \tau)\\
\frac{dy}{dt} &=& - ay \cdot sin( (\omega_1 + c_1) \cdot \tau) \cdot sin(\omega_2)\cdot 2\\
\frac{dz}{dt} &=& - az \cdot sin( (\omega_2 + c_2) \cdot \tau) \cdot cos(\omega_1)\cdot 2\\
\frac{\omega_1}{dt} &=& - c_3\cdot \omega_1\ \ \ \frac{\omega_2}{dt} = -c_4\cdot\omega_2\ \ \ \frac{d\tau}{dt} = 1
\end{eqnarray*}

\textbf{\textit{Quadcopter Control.}} We developed a model that contains the full dynamics of a quadcopter. We use the model to solve control problems by answering reachability questions. A typical set of the differential equations are the following.
\begin{eqnarray*}
\frac{\mathrm{d}\omega_x}{\mathrm{d}t} &=& L\cdot k\cdot (\omega_1^2 - \omega_3^2)(1/I_{xx})-(I_{yy} - I_{zz})\omega_y\omega_z/I_{xx}\\
\frac{\mathrm{d}\omega_y}{\mathrm{d}t} &=& L\cdot k\cdot(\omega_2^2 - \omega_4^2)(1/I_{yy})-(I_{zz} - I_{xx})\omega_x\omega_z/I_{yy}\\
\frac{\mathrm{d}\omega_z}{\mathrm{d}t} &=& b\cdot(\omega_1^2 - \omega_2^2 + \omega_3^2 - \omega_4^2)(1/I_{zz})-(I_{xx} - I_{yy})\omega_x\omega_y/I_{zz}\\
\frac{\mathrm{d}\phi}{\mathrm{d}t} &=& \omega_x + \displaystyle{\frac{\sin\left(\phi\right) \sin\left(\theta\right)}{{\left(\frac{\sin\left(\phi\right)^{2} \cos\left(\theta\right)}{\cos\left(\phi\right)} + \cos\left(\phi\right) \cos\left(\theta\right)\right)} \cos\left(\phi\right)}}\omega_y\\
& &+\displaystyle\frac{\sin\left(\theta\right)}{\frac{\sin\left(\phi\right)^{2} \cos\left(\theta\right)}{\cos\left(\phi\right)} + \cos\left(\phi\right) \cos\left(\theta\right)}\omega_z\\
\frac{\mathrm{d}\theta}{\mathrm{d}t} &=& -(\displaystyle\frac{\sin\left(\phi\right)^{2} \cos\left(\theta\right)}{{\left(\frac{\sin\left(\phi\right)^{2} \cos\left(\theta\right)}{\cos\left(\phi\right)}\omega_y + \cos\left(\phi\right) \cos\left(\theta\right)\right)} \cos\left(\phi\right)^{2}}\\
& &+ \frac{1}{\cos\left(\phi\right)})\omega_y -\displaystyle\frac{\sin\left(\phi\right) \cos\left(\theta\right)}{{\left(\frac{\sin\left(\phi\right)^{2} \cos\left(\theta\right)}{\cos\left(\phi\right)} + \cos\left(\phi\right) \cos\left(\theta\right)\right)} \cos\left(\phi\right)}\omega_z \\
\frac{\mathrm{d}\psi}{\mathrm{d}t} &=& \displaystyle\frac{\sin\left(\phi\right)}{{\left(\frac{\sin\left(\phi\right)^{2} \cos\left(\theta\right)}{\cos\left(\phi\right)} + \cos\left(\phi\right) \cos\left(\theta\right)\right)} \cos\left(\phi\right)}\omega_y\\
& &+ \displaystyle\frac{1}{\frac{\sin\left(\phi\right)^{2} \cos\left(\theta\right)}{\cos\left(\phi\right)} + \cos\left(\phi\right) \cos\left(\theta\right)}\omega_z\\
\frac{\mathrm{d}{xp}}{\mathrm{d}t} &=& (1/m)(\sin(\theta)\sin(\psi)k(\omega_1^2 + \omega_2^2 +\omega_3^2+\omega_4^2) - k\cdot d\cdot{xp})\\
\frac{\mathrm{d}{yp}}{\mathrm{d}t} &=& (1/m)(-\cos(\psi)\sin(\theta)k(\omega_1^2 + \omega_2^2 +\omega_3^2+\omega_4^2) - k\cdot d\cdot{yp})\\
\frac{\mathrm{d}{zp}}{\mathrm{d}t} &=& (1/m)(-g-\cos(\theta)k(\omega_1^2 + \omega_2^2 +\omega_3^2+\omega_4^2) - k\cdot d\cdot{zp}\\
\frac{\mathrm{d}x}{\mathrm{d}t} &=& {xp}, \frac{\mathrm{d}y}{\mathrm{d}t} = {yp}, \frac{\mathrm{d}z}{\mathrm{d}t} = {zp}
\end{eqnarray*}

\textbf{\textit{Prostate Cancer Treatment.}} The Prostate Cancer Treatment model exhibits more nonlinear ODEs. 
\begin{eqnarray*}
\frac{dx}{dt} &=& (\alpha_x
(k_1+(1-k_1)\frac{z}{z+k_2}-\beta_x( (1-k_3)\frac{z}{z+k_4}+k_3))\\
& &- m_1(1-\frac{z}{z_0}))x + c_1 x\\
\frac{dy}{dt} &=& m_1(1-\frac{z}{z_0})x+(\alpha_y (1- d\frac{z}{z_0}) - \beta_y)y+c_2y\\
\frac{dz}{dt} &=& \frac{-z}{\tau} + c_3z\\
\frac{dv}{dt} &=& (\alpha_x
(k_1+(1-k_1)\frac{z}{z+k_2}-\beta_x(k_3+(1-k_3)\frac{z}{z+k_4}))\\
& &- m_1(1-\frac{z}{z_0}))x + c_1 x + m_1(1-\frac{z}{z_0})x+(\alpha_y (1- d\frac{z}{z_0})\\
& &- \beta_y)y+c_2y
\end{eqnarray*}

\begin{figure*}[ht]
\centering
\includegraphics[width=\linewidth]{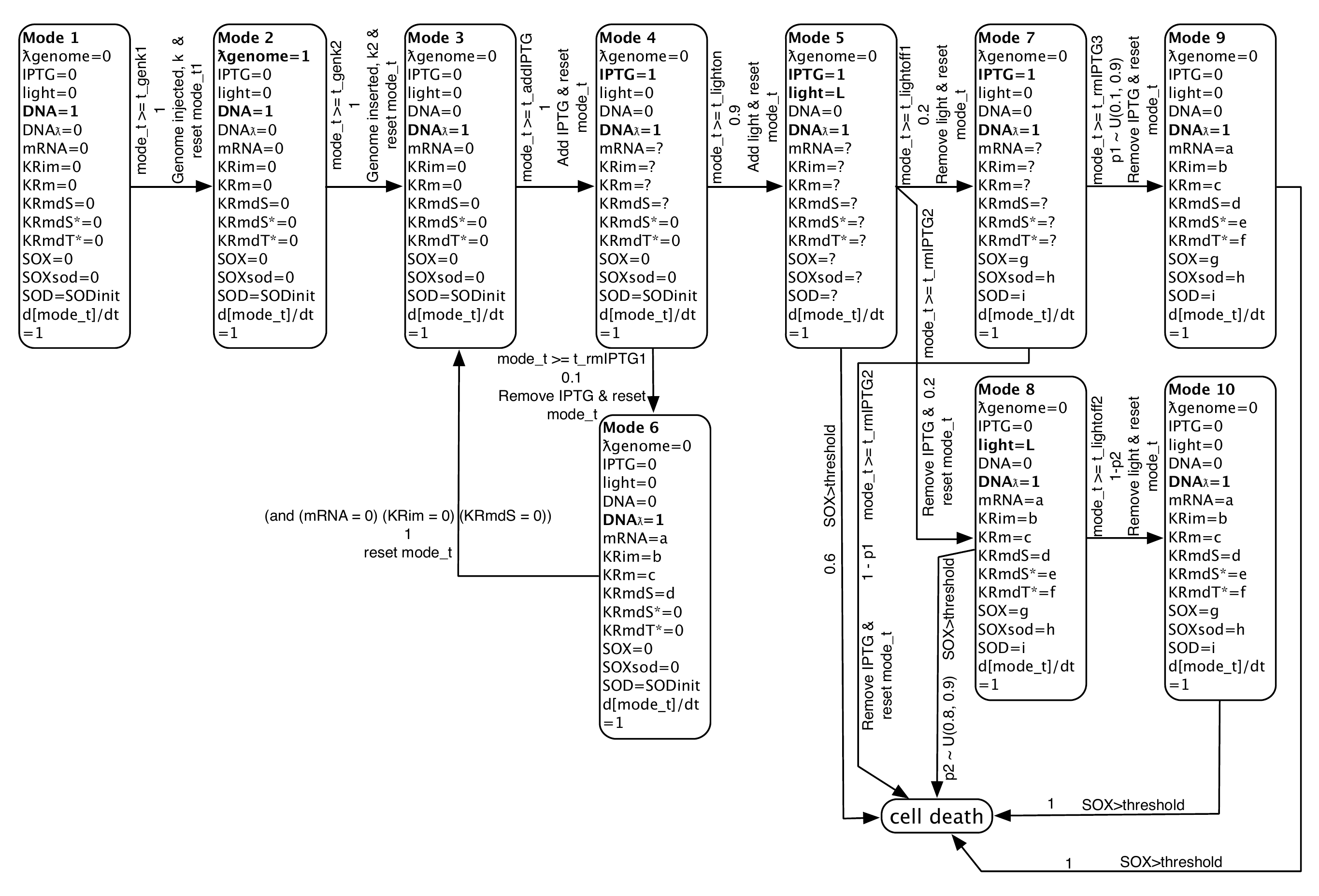}
\caption{A probabilistic hybrid automaton for synthesized phage-based therapy model}
\label{fig:killerred}
\end{figure*}

\textbf{\textit{Synthesized Killerred Model.}} The ODEs missing in Figure \ref{fig:killerred} are as follows.
\begin{eqnarray*}
\frac{\mathrm{d}[mRNA]}{\mathrm{d}t} & =& k_{RNAsyn} \cdot [DNA] - k_{RNAdeg} \cdot [mRNA]\\
\frac{\mathrm{d}[KR_{im}]}{\mathrm{d}t} & =& k_{KR_{im}syn} \cdot [mRNA] - (k_{KR_m} + k_{KR_{im}deg})\\
& & \cdot [KR_{im}]\\
\frac{\mathrm{d}[KR_{mdS}]}{\mathrm{d}t} &=& k_{KR_{m}} \cdot [KR_{im}] - k_{KR_{mdS}deg} \cdot [KR_{mdS}]\\
& &(before\; turning \; on \; the \; light)\\
\frac{\mathrm{d}[KR_{mdS}]}{\mathrm{d}t} &=& k_{KR_{m}} \cdot [KR_{im}] + k_{KR_f} \cdot [KR_{mdS^*}] \\
& &+ k_{KR_{ic}} \cdot [KR_{mdS^*}] + k_{KR_{nrd}} \cdot [KR_{mdT^*}]\\
& &+ k_{KR_{SOXd1}} \cdot [KR_{mdT^* }] - k_{KR_{ex}} \cdot [KR_{mdS}]\\
& &- k_{KR_{mdS}deg} \cdot [KR_{mdS}]\;\;\; (after\; adding\; light)\\
\frac{\mathrm{d}[KR_{mdS^*}]}{\mathrm{d}t} &=& k_{KR_{ex}} \cdot [KR_{mdS}] - k_{KR_f } \cdot [KR_{mdS^*}] \\
& &-k_{KR_{ic}} \cdot [KR_{mdS^*}] - k_{KR_{isc}} \cdot [KR_{mdS^*}]\\
& &- k_{KR_{mdS^*}deg} \cdot [KR_{mdS^*}]\\
\frac{\mathrm{d}[KR_{mdT^*}]}{\mathrm{d}t} &=& k_{KR_{isc}} \cdot [KR_{mdS^*}] - k_{KR_{nrd}} \cdot [KR_{mdT^*}]\\
& &- k_{KR_{SOXd1}} \cdot [KR_{mdT^*}]\\
& &- k_{KR_{SOXd2}} \cdot [KR_{mdT^*}]\\
& &- k_{KR_{mdT^*}deg} \cdot [KR_{mdT^*}]
\end{eqnarray*}
\begin{eqnarray*}
\frac{\mathrm{d}[SOX]}{\mathrm{d}t} &=& k_{KR_{SOXd1}} \cdot [KR_{mdT^*}] + k_{KR_{SOXd2}} \cdot [KR_{mdT^*}]\\
& &- \frac{\mathrm{d}[SOX_{sod}]}{\mathrm{d}t}\\
\frac{\mathrm{d}[SOX_{sod}]}{\mathrm{d}t} &=& k_{SOD} \cdot V_{maxSOD} \cdot \frac{[SOX]}{K_m + [SOX]}
\end{eqnarray*}

\vspace{.5cm}
\section{Experimental results for additional benchmarks}\label{apndx:exp}
\begin{center}
\begin{table*}[ht]
\label{table:exp}
\captionsetup{font=scriptsize}
\centering
\begin{tabular}{c|c|c|c|c|c|c|c|c|c|c|c}
\hline
Benchmark & \#Ms & K & \#ODEs & \#Vs & \#RVs & $\delta$ & Est\_P & \#S\_S & \#T\_S & A\_T(s) & T\_T(s) \\ \hline
BBK1 & 1 & 1 & 2 & 14 & 3 & 0.001 & 0.754 & 5372 & 7126 & 0.086 & 612.836 \\ \hline
BBK5 & 1 & 5 & 2 & 38 & 3 & 0.001 & 0.059 & 209 & 3628 & 0.253 & 917.884 \\ \hline
BBwDv1 & 2 & 2 & 4 & 20 & 4 & 0.001 & 0.208 & 2206 & 10919 & 0.080 & 873.522 \\ \hline
BBwDv2K2 & 2 & 2 & 4 & 20 & 3 & 0.001 & 0.845 & 7330 & 8669 & 0.209 & 1811.821 \\ \hline
BBwDv2K8 & 2 & 8 & 4 & 56 & 3 & 0.001 & 0.207 & 2259 & 10901 & 0.858 & 9353.058 \\ \hline
Tld & 2 & 7 & 2 & 33 & 4 & 0.001 & 0.996 & 227 & 227 & 0.213 & 48.351 \\ \hline
Ted & 2 & 7 & 4 & 50 & 4 & 0.001 & 0.996 & 227 & 227 & 12.839 & 2914.448 \\ \hline
DTldK3 & 2 & 3 & 4 & 26 & 2 & 0.001 & 0.996 & 227 & 227 & 0.382 & 86.714 \\ \hline
DTldK5 & 2 & 5 & 4 & 38 & 2 & 0.001 & 0.161 & 1442 & 8961 & 0.280 & 2509.078 \\ \hline
W4mv1 & 4 & 3 & 8 & 26 & 6 & 0.001 & 0.381 & 5953 & 15639 & 0.238 & 3722.082 \\ \hline
W4mv2K3 & 4 & 3 & 8 & 26 & 6 & 0.001 & 0.996 & 227 & 227 & 0.673 & 152.771 \\ \hline
W4mv2K7 & 4 & 7 & 8 & 50 & 6 & 0.001 & 0.004 & 0 & 227 & 0.120 & 27.240 \\ \hline
DWK1 & 2 & 1 & 4 & 14 & 5 & 0.001 & 0.996 & 227 & 227 & 0.171 & 38.817 \\ \hline
DWK3 & 2 & 3 & 4 & 26 & 5 & 0.001 & 0.996 & 227 & 227 & 0.215 & 48.806 \\ \hline
DWK9 & 2 & 9 & 4 & 62 & 5 & 0.001 & 0.996 & 227 & 227 & 5.144 & 1167.688 \\ \hline
Que & 3 & 2 & 3 & 13 & 4 & 0.001 & 0.228 & 2662 & 11677 & 0.095 & 1109.315 \\ \hline
3dOsc & 3 & 2 & 18 & 48 & 2 & 0.001 & 0.996 & 227 & 227 & 8.273 & 1877.969 \\ \hline
QuadC & 1 & 0 & 14 & 44 & 6 & 0.001 & 0.996 & 227 & 227 & 825.641 & 187420.507 \\ \hline
ExPHA01 & 2 & 2 & 4 & 20 & 2 & 0.001 & 0.524 & 345 & 658 & 5.01 & 3295.82 \\ \hline
ExPHA02 &  2  &  3  & 2  & 17 & 1  &  0.001 &  0.900  & 5361 & 5953  & 0.0004   & 2.35  \\ \hline
KRk5 & 6   &  5  & 84  &194  & 2  &  0.001 &  0.544 & 8946   & 16457     &  0.122  & 2015.64  \\ \hline
KRk6 & 8   &  6  &112   &224  &  6 &  0.001 &  0.246  & 2032   & 8263     & 1.385   & 11444.22  \\ \hline
KRk7 & 10   &  7  & 150  &271  & 6  &  0.001 & 0.096 & 558   & 5795     & 16.275   & 94311.18  \\ \hline
KRk8 &  7  &  8  & 105  &303  & 6  &  0.001 &   0.004       &  0  & 227     & 0.003   & 0.58  \\ \hline
KRk9 &  9  &  9  & 135  & 335 & 6  &  0.001 &   0.004       & 0   & 227     & 0.015   &3.43   \\ \hline
KRk10 & 11   &  10  & 165  & 367 & 6  &  0.001 &  0.004        &0   &  227    & 0.026   &  5.92 \\ \hline
\end{tabular}
\caption {\#Ms = number of modes, K indicates the unfolding steps, \#ODEs = number of ODEs in the model, \#Vs = number of total variables in the unfolded formulae, \#RVs = number of random variables in the model, $\delta$ = precision used in {\it dReach}, \#S\_S = number of $\delta$-sat samples , \#T\_S = total number of samples, Est\_P = estimated maximum posterior probability,  A\_T(s) = average CPU time of each sample in seconds, and T\_T(s) = total CPU time for all samples in seconds.}
\label{table:additonalexp}
\end{table*}
\end{center}

\begin{table*}[th!]
\captionsetup{font=scriptsize}
\centering
    \begin{tabular}{|c|c|c|c|c|c|c|c|c|c|c|}
    \hline
    $t_{lighton}$ & 1 & 2 & 3 & 4 & 5 & 6 & 7 & 8 & 9 & 10 \\ \hline
    $t_{tot}$ & 16  & 17.2     &  18.5   & 20   & 21.3 & 22.7 & 23.5 & 24.1 & 25 & 30    \\ \hline
    \end{tabular}
    \caption{The relation between the time to turn on the light after adding IPTG and the total time to kill bacteria cells ($k=5$). }
    \label{table:kr02}
\end{table*}

\begin{table*}[th!]
\captionsetup{font=scriptsize}
\centering
    \begin{tabular}{|c|c|c|c|c|c|c|c|c|c|c|}
    \hline
    $t_{lightoff1}$ & 1 & 2 & 3 & 4 & 5 & 6 & 7 & 8 & 9 & 10 \\ \hline
    $kill bacteria cells$ & Failed  & Failed    &  Failed   & Succ   & Succ & Succ & Succ & Succ & Succ & Succ    \\ \hline
    \end{tabular}
    \caption{The impact of the time duration that the cells are exposed to light ($k=6$).}
    \label{table:kr03}
\end{table*}

The table \ref{table:additonalexp} shows the results of experiments. To further demonstrate the feasibility of {\it SReach}, these experiments were conducted with the sequential version of {\it SReach} on a machine with 2.9GHz Intel Core i7 processor and 8GB RAM, running OS X 10.9.2. In our experiments we used $0.001$ as the precision for the $\delta$-decision problem; and Bayesian sequential estimation with $0.01$ half-interval width, coverage probability $0.99$, and uniform prior ($\alpha = \beta = 1$). In the following table, BB refers to the bouncing ball models, Tld the thermostat model with linear temperature decrease, Ted the thermostat model with exponential decrease, DT the dual thermostat models, W the watertank models, DW the dual watertank models, Que the model for queuing system which has both nonlinear functions and nondeterministic jumps, 3dOsc the model for 3d oscillator, and QuadC the model for quadcopter stabilization control. Following these hybrid systems with parametric uncertainty, we also consider two example PHAs - ExPHA01and EXPHA02, and PHAs with additional randomness - KR our killerred models.

\end{document}